\documentclass[10pt, conference, letterpaper]{IEEEtran}
\IEEEoverridecommandlockouts
\usepackage{cite}
\usepackage{amsmath,amssymb,amsfonts}
\usepackage{algorithmic}
\usepackage{graphicx}
\usepackage{textcomp}
\usepackage{xcolor}
\usepackage{indentfirst}
\usepackage{multirow}
\usepackage{makecell}
\usepackage{enumitem}
\usepackage{afterpage} 
\usepackage{lipsum} 
\usepackage{float} 
\usepackage{multicol}
\usepackage{stfloats}
\def\BibTeX{{\rm B\kern-.05em{\sc i\kern-.025em b}\kern-.08em
    T\kern-.1667em\lower.7ex\hbox{E}\kern-.125emX}}

\begin{document}

\title{Computation Pre-Offloading for MEC-Enabled Vehicular Networks via Trajectory Prediction
}

\author{
\IEEEauthorblockN{Ting Zhang\IEEEauthorrefmark{1}, Bo Yang\IEEEauthorrefmark{2}, Zhiwen Yu\IEEEauthorrefmark{2},  Xuelin Cao\IEEEauthorrefmark{3}, George C. Alexandropoulos\IEEEauthorrefmark{4}, Yan Zhang\IEEEauthorrefmark{5}, 
and Chau Yuen\IEEEauthorrefmark{6}} 

\IEEEauthorblockA{\IEEEauthorrefmark{1}School of Software, Northwestern Polytechnical
University, Xi'an, Shaanxi, 710129, China} 
\IEEEauthorblockA{\IEEEauthorrefmark{2}School of Computer Science, Northwestern Polytechnical
University, Xi'an, Shaanxi, 710129, China} 
\IEEEauthorblockA{\IEEEauthorrefmark{3}School of Cyber Engineering, Xidian University, Xi'an, Shaanxi, 710071, China}
\IEEEauthorblockA{\IEEEauthorrefmark{4}Department of Informatics and Telecommunications, National and Kapodistrian University of Athens, 15784 Athens, Greece}
\IEEEauthorblockA{\IEEEauthorrefmark{5}Department of Informatics, University of Oslo, 0316 Oslo, Norway}
\IEEEauthorblockA{\IEEEauthorrefmark{6}School of Electrical and Electronics Engineering, Nanyang Technological University, Singapore}

}
\maketitle
\begin{abstract}
Task offloading is of paramount importance to efficiently orchestrate vehicular wireless networks, necessitating the availability of information regarding the current network status and computational resources. However, due to the mobility of the vehicles and the limited computational resources for performing task offloading in near-real-time, such schemes may require high latency, thus, become even infeasible. To address this issue, in this paper, we present a Trajectory Prediction-based Pre-offloading Decision (TPPD) algorithm for analyzing the historical trajectories of vehicles to predict their future coordinates, thereby allowing for computational resource allocation in advance. We first utilize the Long Short-Term Memory (LSTM) network model to predict each vehicle's movement trajectory. Then, based on the task requirements and the predicted trajectories, we devise a dynamic resource allocation algorithm using a Double Deep Q-Network (DDQN) that enables the edge server to minimize task processing delay, while ensuring effective utilization of the available computational resources. Our simulation results verify the effectiveness of the proposed approach, showcasing that, as compared with traditional real-time task offloading strategies, the proposed TPPD algorithm significantly reduces task processing delay while improving resource utilization.
\end{abstract}
\begin{IEEEkeywords}
Internet of vehicles, computation offloading, long short-term memory, double deep Q-network.
\end{IEEEkeywords}

\section{Introduction}

With the rapid development of mobile/multi-access edge computing (MEC) technology, vehicles can now utilize the computing power of edge servers to efficiently handle tasks. However, due to the mobility of vehicles, the available edge servers in the vicinity are constantly changing. At the same time, individual tasks vary in terms of computational requirements, latency sensitivity, data size, etc., so choosing an appropriate offloading strategy becomes a key issue. 
Based on different optimization objectives, some researchers proposed feasible approaches for vehicle task offloading strategies. For example, the authors of \cite{ref1} proposed a deep Q-network-based resource allocation algorithm that improves the system throughput while satisfying the service delay requirement. Another study \cite{ref2} introduced a deep Q-learning algorithm with the weighted sum of delay and energy consumption as the optimization objective. In this scenario, individual mobile devices make offloading decisions without knowing the task model. 
Markov decision process (MDP) is used in \cite{ref3} to model the queuing state of mobile devices and MEC servers, and then a joint user association and resource allocation algorithm is given by combining MDP and matching theory. In \cite{ref4}, a multi-task learning-based feed-forward neural network (MTFNN) model is designed and trained to jointly optimize offloading decisions and computational resource allocation. Additionally, a probabilistic computation offloading algorithm is proposed in \cite{ref5} to solve the cooperative computation offloading problem, which is based on a convex framework known as the alternating direction multiplier method to achieve the optimal allocation probability iteratively. The authors of \cite{ref8} propose a joint system load optimization approach based on the discrete binary particle swarm optimization algorithm to optimize task offloading and reduce latency and energy consumption. In \cite{ref9}, the author proposes a deep reinforcement learning (DRL) approach to achieve the minimization of total energy consumption in a binary offload scenario with MEC servers and multiple smart devices. The authors of \cite{ref13} propose deploying Reconfigurable Intelligent Computing Surfaces (RICS) in the Telematics network to leverage the computational power of its metamaterials to maximise safety-based autonomous driving tasks. In \cite{ref15}, the author introduces drones as an additional access point. A multi-intelligence reinforcement learning framework is utilised to maximise user-drone association by jointly optimising drone trajectories and user association metrics under quality of service constraints.

\begin{figure}[t]
    \centering
    \includegraphics[width=3.35in, height=2.55in]{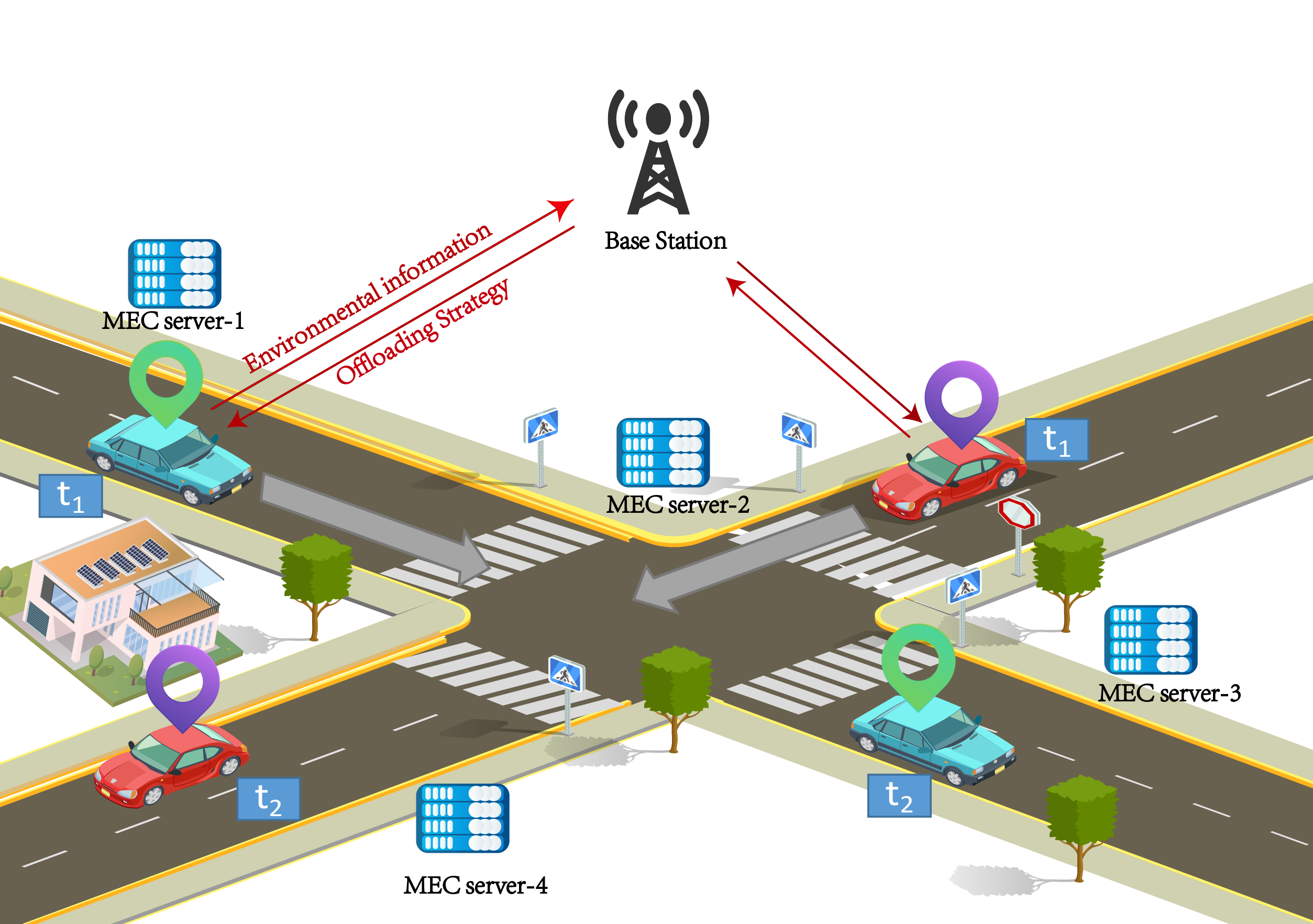}
    \caption{The considered scenario of multiple network-connected vehicles uploading environmental information to the network and performing task offloading in nearby MEC servers.}
    \label{changjing}
\end{figure}

Note that solving task offloading optimization problems is usually time-consuming, so after obtaining the offloading decision,  the vehicle may arrive at a totally different MEC environment, as illustrated in Fig.~\ref{changjing}. In this case, for the blue vehicle, the offloading decision made under the environment at the time $t_1$ (e.g., the MEC server 1 is available) may become invalid at the time $t_2$ (e.g., the MEC server 3 is available), where $\triangle t\!=\!t_2-t_1$ indicates the time cost for obtaining the offloading decision. 
Therefore, it is necessary to make an offloading decision at the time $t_1$ in advance for the time $t_2$. However, none of the works mentioned above consider this frequently occurring situation. These works neither take into account the mobility of the vehicle nor the varying demands of the tasks in the offloading decisions and resource allocation, thereby reducing the system's resource utilization. 

Although the computation offloading problem has been extensively explored over the past few years, appropriate offloading decision-making still remains a challenge in the dynamical vehicular MEC system. This is mainly due to the constant change of the available MEC servers, which largely depends on the movement trajectory of vehicles. Suppose that a fast-moving vehicle would like to make an offloading decision for its task. 
Therefore, it is still crucial to study the joint motion trajectory prediction for task-offloading decision-making in advance.  
To address this issue, this paper introduces a 
trajectory prediction-based pre-offloading decision (TPPD) algorithm for multi-vehicle and multi-server scenarios, where we aim at making optimal offloading strategies in advance based on trajectory prediction. 
 Numerical results demonstrate that our proposed scheme significantly outperforms existing methods, including DDQN \cite{ref10}, DQN \cite{ref11}, exhaustive algorithms, and traditional offloading schemes, in terms of time delay and power consumption.
 
\section{System Model}
\subsection{Network Scenario}
This paper focuses on a multi-vehicle multi-server scenario, where there are $N$ vehicles, represented as \({\cal U} \!=\! \{U_1, U_2,  \ldots, U_N\}\), and $M$ MEC servers that are distributed throughout the area, denoted as \({\cal E} \!=\! \{E_1, E_2, \ldots, E_M\}\), each with the computational resources denoted as \(F_i\), where \(i\) denotes different MEC servers. The total time is divided into multiple time slots with equal length, denoted as ${\cal T} = \{ t_0, t_1, \ldots, t_m \}$. In each time slot, each vehicle within the coverage area of the MEC servers can select one of the servers to offload the tasks via wireless links, according to the trajectory prediction and computational resource allocation of MEC servers, as shown in Fig.~\ref{changjing}. Once the wireless link is established, the vehicle can send the computation task to the selected MEC server, which helps with the computation task processing and returns the result to the vehicle.

Assuming that the tasks cannot be further split, we introduce a quaternion to denote the computational task of the $i$-th vehicle \(U_i\) in the $j$-th time slot \(t_j\), i.e., \(task_{j,i}\!=\!(l_{j, i}, c_{j, i}, \nu_{j, i}, Z_{j, i})\), where \(l_{j, i}\) denotes the amount of data in the computational task, \(c_{j, i}\) denotes the amount of computation needed to process the task, \(\nu_{j, i}\) denotes the tolerable latency of the task, and \(Z_{j, i}\) denotes the priority of the task. The historical trajectory of the vehicle $U_i$ up to the time moment \(t\) can be represented as \({\cal P}_{t,i} = \{ P_{1,i}, P_{2, i}, \ldots, P_{t-1,i} \}\).
\subsection{Local Computation Model}
In practice, some tasks with small data volume, small tolerable latency, and sometimes privacy requirements are prone to be executed locally, thereby guaranteeing the timeliness and data privacy of the tasks. In this case, the local execution latency of \(task_{j,i}\) is calculated by:
\[
t^l_{j,i} = \frac{c_{j,i}}{F^l_{i}}, \tag{1}
\]
where \(F^l_{i}\) denotes the computing resources of the vehicle $U_i$.
\subsection{Communications Model}
We consider each vehicle communicates with the MEC server via orthogonal frequency-division multiplexing (OFDM), where the total frequency bandwidth (denoted as $B$) is assumed to be divided into \(N\) subchannels with an equal bandwidth. For the convenience of research, we consider that the channel state information (CSI) between the vehicle ($U_i, 1\leq i\leq N$) and the MEC server ($E_k, 1\leq k\leq M$) does not change within the same time slot. Then the transmission rate of the vehicle $U_i$ at the $j+1$ time slot can be expressed as
\[
R_{j+1,i,k} = \frac{B}{N}\log_2\left(1+\frac{p_i h_{j+1,i,k}}{\sigma^2}\right),\tag{2}
\]
where 
\(B\) is the total bandwidth of the wireless channel, which is divided into \(N\) subchannels with equal bandwidth. \(p_i\) denotes the transmission power of the vehicle $U_i$, \(\sigma^2\) denotes Gaussian white noise. \(h_{j+1,i,k}\) is the channel gain between the vehicle $U_i$ and MEC server \(E_k\), which is given by \cite{ref6}
\[
h_{j+1,i,k} = h_0\left(\frac{d_0}{d_{j+1,i,k}}\right)^r, \tag{3}
\]
where \(r\) denotes the path loss exponent, which is generally assumed to be 2. \(d_0\) denotes the reference distance between the vehicle and the edge server, \(d_{j+1, i, k}\) is the actual distance between $U_i$ and $E_k$ at the time slot $t_{j+1}$, and \(h_0\) is the path loss constant.

\subsection{Computation Offloading Model}
Computation offloading is usually applied when a vehicle node decides to offload a task to an edge server to achieve more computational resources to help with task execution. The total latency of task processing at the edge server consists of three parts: the latency of offloading the task to the edge server over the wireless link, the computation latency raised at the edge server, and the latency of transmitting the results of the computation back to the vehicle, which can be ignored due to the small size of the results. Therefore, the total latency of the task offloaded to the edge server is given by
\[
t^{\text{off}}_{j+1,i,k} = t^{\text{trans}}_{j+1,i,k} + t^{\text{edge}}_{j+1,i,k}. \tag{4}
\]
where \( t^{\text{trans}}_{j+1, i,k} \) denotes the time spent offloading the task to the edge server, i.e., \( t^{\text{trans}}_{j+1, i,k} = \frac{l_{j, i}}{v_{j+1, i,k}} \). \( t^{\text{edge}}_{j+1, i,k} \) indicates the computational latency of the task at the server, i.e., \( t^{\text{edge}}_{j+1, i,k} = \frac{c_{j, i}}{\omega_{j+1, i,k} F_{j+1,k}} \).
Here, \( \omega_{j+1, i,k} \) denotes the percentage of computational resources allocated by the edge server, which satisfies the following constraint:\(\sum_{i=1}^{N} \omega_{j+1,i,k} \leq 1.\)
\begin{figure*}[ht]
    \centering
    \includegraphics[width=1\textwidth]{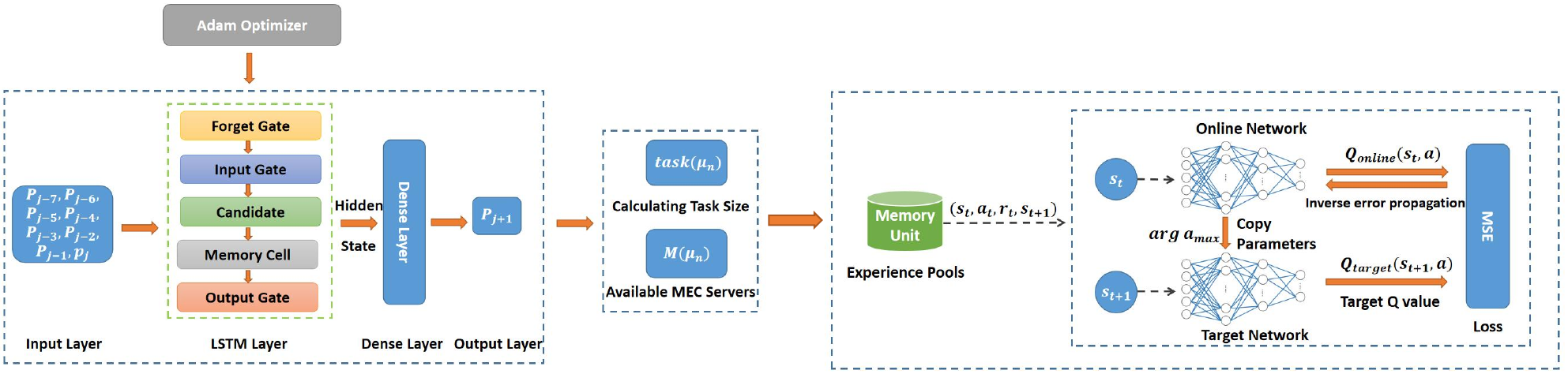}
    \caption{The proposed TPPD algorithm architecture.}
    \label{fig_framework}
\end{figure*}
\subsection{Problem Formulation}
Our goal is to predict the trajectories of the joint vehicles under resource constraints, task maximum up-to-date time constraints, and thus minimize the overall delay. Thus, the optimization problem can be formulated as:
\begin{align}
\min_{x_i,y_{i,k},\omega_{u,k}} & \quad \sum_{i=1}^{N}\sum_{k=1}^{K}\sum_{\omega_{u,k}=1}^{N} \left[ x_i y_{i,k} t^{\text{off}}_{j+1,i,k} + (1-x_i) t^l_{j,i} \right] \notag\\
\text{s.t.} \quad C_1: & \quad \eta_{x_i} \in \{0, 1\}, \tag{5-a} \\
C_2: & \quad y_{i,k} \in \{0, 1\}, \ y_{i,1} \!+\! y_{i,2} \!+\! \ldots \!+\! y_{i,K} \!=\! 1, \tag{5-b} \\
C_3: & \quad \sum_{t=1}^{N} \omega_{t,k} = 1, \ k \in \{1, 2, \ldots, M\}, \tag{5-c} \\
C_4: & \quad C < \tau. \tag{5-d}
\end{align}\par
If the task chooses to be offloaded to a server, we have \(x_i=1\); otherwise, \(x_i=0\). \(y_{i,k}\) indicates which server it chooses to offload to. \(\omega_{i,k}\) indicates the percentage of computational resources allocated by the edge server.
\section{Proposed PRE-TASK OFFLOADING Approach}
In this Section, we propose the TPPD algorithm to solve the formulated problem. We take the position information of eight consecutive moments in the vehicle history as the input to the whole network and finally output the offloading strategy for the next position through the TPPD network.\par
As shown in Fig. \ref{fig_framework}, the proposed TPPD algorithm is mainly divided into two parts. The first part uses long short-term memory (LSTM) to predict the position information of the next moment based on the historical trajectory of the vehicle. The other part is based on the results obtained in the first step, a DDQN-based algorithm is designed to optimize the task offloading strategy. In this context, we can decompose the complex problem into two sub-problems, i.e., trajectory prediction and pre-offloading strategy design.
\subsection{Vehicle Trajectory Prediction}\label{AA}
Trajectory prediction is a typical time-series problem, where the current position of a vehicle depends not only on the recent state but is also affected by the state in the past period of time. LSTM, by its design, can efficiently capture and retain these long-time dependencies. LSTM is capable of learning and recognizing complex dynamic patterns and behaviors, such as the acceleration, deceleration, and steering behaviors of a vehicle under different traffic conditions. Having a powerful remember-and-forget mechanism allows us to predict the position information at the next moment using the vehicle's historical continuous trajectory information, which includes the vehicle's latitude and longitude information at each moment.\par
\textit{1) Data Pre-processing:} Based on the differences in the trajectory data of the vehicles, to prevent the instability of the values caused by some values being too large or too small, we first preprocess the raw data for data cleaning to deal with missing values and outliers. We normalize the collected data to an interval $[0, 1]$ using the Min-Max normalization method to maintain the distribution pattern of the data. The normalized values are calculated as follows:
\[
D_{\text{norm}} = \frac{D - D_{\text{min}}}{D_{\text{max}} - D_{\text{min}}} \tag{6}
\]
where \(D\) is the original data in the dataset, \(D_{\text{min}}\) is the minimum value in the dataset, \(D_{\text{max}}\) is the maximum value in the dataset, and \(D_{\text{norm}}\) is the normalized data.\par
\textit{2) LSTM-based Trajectory Prediction:} The position information of the vehicle at eight consecutive moments in its history is used as input to the model, and then to the LSTM layer, which is the core of the network, to process the time-series data. The LSTM processes and stores the time-step dependencies between time steps using memory cells and a gate mechanism. These mechanisms enable the LSTM to capture and retain the long-term dependencies and complex dynamic patterns within the trajectory data.\par
This layer outputs the hidden state of each time step and passes it on to the next layer. First, a forgetting gate determines how much of the memory cell state from the previous time step should be retained and forgotten. Then, the input gate combines the new inputs to decide how much of the new information from the current time step needs to be written to the memory cell state, followed by generating new candidate memory content and updating the memory cell state. Finally, the output gate decides the output of the current memory cell state. The model uses mean square error (MSE) as the loss function and Adam optimizer for optimization.\par
Subsequently, the fully connected layer receives the hidden state output from the LSTM and maps it to the output layer to get the position coordinates of the next moment.\par
\subsection{MEC Server Filtering}
In the MEC-enabled vehicular networks, the vehicles perform task offloading while moving with high speeds. Frequent server switching leads to an increase of signaling overhead, which introduce additional overall latency and thus affect network efficiency. To address this issue, we prepare the edge server at the next location in advance based on the predicted location at the next moment. Based on the signal strength and service range of the MEC servers, we prioritize the edge servers, which ensures seamless transfer of tasks during the switchover, reduces interruptions, and improves the offloading efficiency and system stability.\par
The service range of each MEC server is given by \(d^{s}_{\text{max}}\) and the transmission rage of the vehicle is \(d^{v}_{\text{max}}\). The maximum effective communication distance between a vehicle and a server is \(r\). The location information of the vehicle and the edge server is expressed in latitude and longitude. To calculate the distance between the two more accurately, the distance is calculated using Haversine's formula. Assuming two positions are given as \((\varphi_1, \lambda_1)\) and \((\varphi_2, \lambda_2)\), we have
\[
a = \sin^2 \left( \frac{\Delta \varphi}{2} \right) + \cos(\varphi_1) \cos(\varphi_2) \sin^2 \left( \frac{\Delta \lambda}{2} \right), \tag{7}
\]
\[
c = 2 \operatorname{atan2} \left( \sqrt{a}, \sqrt{1-a} \right), \tag{8}
\]
\[
d = R  c, \tag{9}
\]
where
\(\varphi_1, \varphi_2\) are the latitudes of the two points (in radians), 
\(\lambda_1, \lambda_2\) are the longitudes of the two points (in radians), 
\(R\) is the Earth's radius with the mean radius of 6371 km.\par
When the distance between a vehicle and a server is within the communication range of the vehicle and the service range of the server, i.e., $d \leq {\rm min} \{d^{s}_{\text{max}}, d^{v}_{\text{max}} \}$, the MEC servers that meet the condition are considered to be available and the remaining MEC servers can be
excluded. 
\subsection{Pre-Offloading Strategy}
In a multi-vehicle multi-server scenario, based on the trajectory prediction results, we can predict the vehicle's moving path and location in the future. Thus, the task offloading strategy can be obtained in advance. However, there are multiple selectable edge servers in the intersection of two location service ranges, and the offloading strategy is achieved to select the optimal MEC server. Due to the dynamics of vehicle locations and network conditions, we propose a DDQN-based offloading algorithm, which obtains the offloading strategy for each task and fully utilizes the limited resources of the server to handle various tasks.\par
\textit{1) Priority Setting:} Considering that tasks have different features and requirements, such as collision detection, emergency braking, and other tasks that need to be processed immediately. To ensure the timeliness of tasks, we design a comprehensive scoring model that integrates the priority, importance, and required resources of in-vehicle tasks.\par
In order to eliminate the influence of the magnitudes between the three indicators, the data need to be standardized. The data are transformed into the same magnitude to retain the distributional characteristics while avoiding the negative impact of magnitude differences on the results. In this paper, we use deviation standardization to scale the values of the original three features to the same range \([0, 1]\) and assign different weights to calculate the importance of the task. Let \(z_{j,i,f}\) denote the priority factor of a task, the normalized data is expressed as
\[
z'_{j,i,f} = \frac{z_{j,i,f} - z^{\min}_{j,i,f}}{z^{\max}_{j,i,f} - z^{\min}_{j,i,f}}, \tag{10}
\]
where \( j \in \{1, 2, 3\} \), \(z^{\min}_{j, i,f}\) and \(z^{\max}_{j, i,f}\) indicate the minimum and maximum values of the three influences, respectively.\par
After standardization, the deviation based on different weights is combined to evaluate the task priority, which is calculated as
\[
Z_{j,i} = \alpha z'_{j,i,1} + \beta z'_{j,i,2} + \lambda z'_{j,i,3}, \tag{11}
\]
where \(\alpha\), \(\beta\), and \(\lambda\) denote the weight values and the weights need to satisfy \(\alpha + \beta + \lambda = 1\). When different tasks are chosen to be offloaded to the same edge server, resources are allocated in equal proportions according to their priorities.
\par
\textit{2) DDQN-based Task Offloading Strategy:} First, we set the intersection of available servers and the tasks at the current moment as the state space.
\[
S_n = \left( \text{task}(\mu_n), M(\mu_n) \right). \tag{12}
\]
The action space, i.e., the set of offloading policies, is then  defined as
\[
A_n = \left\{ 
\begin{matrix}
\alpha_1, \alpha_2, \ldots, \alpha_n \\
\beta_1, \beta_2, \ldots, \beta_n 
\end{matrix} 
\right\}, \tag{13}
\]
\[
\alpha_i = \{0, 1\}, \tag{14}
\]
\[
\beta_i = \{1, 2, \ldots, K\}. \tag{15}
\]\par
When \(\alpha_i = 0\), it means that the vehicle chooses local execution. When \(\beta_i = k\), it means that the vehicle offloads the task to the edge server \(k\). When there is a choice of the same edge server, the allocation of the available resources of the server is done based on priority. After allocating the resources, the predetermined end time of each task and the percentage of server resources used are monitored in real-time. When the task is finished, the computational resources allocated to these vehicles will be released. \par
The overall reward is jointly determined by the state and action before and after taking the corresponding action. It takes into account the effect of the strategy on the overall delay with the aim of minimizing the overall delay. DDQN combines double Q-learning and DQN and is also divided into two phases. The first is the experience storage phase. An experience replay memory is initialized, and the current state is obtained. The agent chooses one of the actions according to the \(\epsilon\)-greedy strategy and performs it in the environment to get a certain reward. The environment then transitions to the next state \(s_{t+1}\). By continuously interacting with the environment, the agent stores its experiences, including the current state, the action taken, the reward received from the action, and the next state, i.e., \((s_t, a, r, s_{t+1})\). This process continues until the capacity of the memory bank is reached, at which point learning begins. When new experiences arrive, the oldest experiences are replaced.\par
In the learning phase, a batch of experience samples are randomly selected from the memory bank, and the following steps are performed: \((s_t)\) is used as input to the online network to get the value of \(Q_{\text{online}}(s_t, a_t)\). \((s_{t+1})\) is input into the target network to get \(Q_{\text{target}}(s_{t+1}, A)\), where \(A\) is the action with the maximum value obtained from \(Q_{\text{target}}\). \(Q_{\text{online}}\) is used as the predicted value of the network, and \(r_{t} + \gamma \times Q_{\text{target}}(s_{t+1}, A)\) is used as the actual value of the network. The error between \(Q_{\text{online}}\) and \(Q_{\text{target}}\) is computed using the mean square error loss function (MSE Loss). The parameters of the Online network are updated by backpropagation to minimize the error. The parameters of the online network are also copied to the target network after a fixed number of learning steps to maintain the stability of the target network.\par
\section{Experimental and Discussion}
We combine GPS data \cite{ref7} as our dataset with open base station information from the OpenCelliD website. There are 6 MEC servers and 4 car devices in the considered scenario, with offloading decisions made at fixed intervals.
\subsection{Trajectory Prediction}
\begin{figure}[t]
    \centering
    \includegraphics[width=0.525\textwidth]{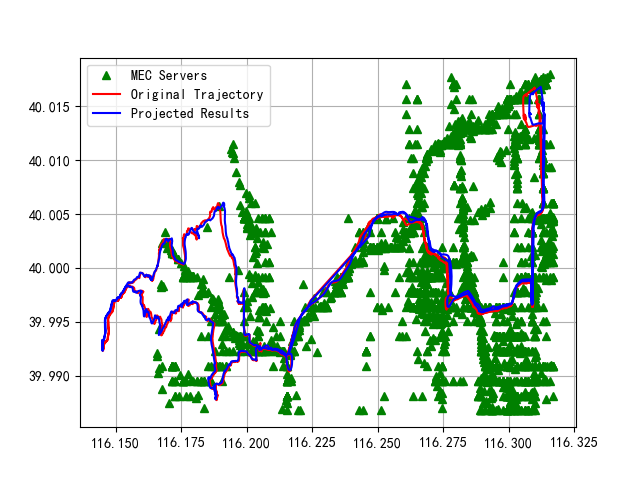}
    \caption{Prediction results of the two-layer LSTM network model.}
    \label{fig:galaxy}
\end{figure}

We filtered a user's movement trajectory for 7082 moments over 12 hours from the dataset and trained it using a two-layer LSTM network model. The prediction results we obtained are shown in Fig. \ref{fig:galaxy}.
\par
To more comprehensively assess the effectiveness and accuracy of the algorithmic model in trajectory prediction, the mean square error (MSE), the root mean square error (RMSE), the mean absolute error (MAE) and the accuracy rate \((A_c)\) are used as evaluation metrics, which are computed as follows:
\[
E_{\text{MAE}} = \frac{1}{n} \sum_{t=1}^{n} \left| \theta_{t+1} - \hat{\theta}_{t+1} \right|, \tag{17}
\]
\[
E_{\text{MSE}} = \frac{1}{n} \sum_{t=1}^{n} \left(\theta_{t+1} - \hat{\theta}_{t+1} \right)^2, \tag{18}
\]
\[
E_{\text{RMSE}} = \sqrt{\frac{1}{n} \sum_{t=1}^{n} \left( \theta_{t+1} - \hat{\theta}_{t+1} \right)^2}, \tag{19}
\]
\[
A_C = 1 - E_{\text{RMSE}}, \tag{20}
\]
where \(n\) denotes the amount of data predicted for the dataset, \(\theta_{t+1}\) denotes the true position at moment \(t+1\), and \(\hat{\theta}_{t+1}\) denotes the predicted position at moment \(t+1\). The results of the evaluation metrics predicted by the different trajectory datasets are shown in Table \ref{tab1}.\par
\renewcommand{\arraystretch}{1.5}
\begin{table}[h]
\caption{Movement trajectory prediction error}
\begin{center}
\resizebox{\columnwidth}{!}{
\begin{tabular}{|c|c|c|c|c|}
\hline
\multirow{2}{*}{\textbf{Datasets}} & \multicolumn{4}{c|}{\textbf{Evaluation Indicators}} \\ \cline{2-5}& $E_{MAE}$ & $E_{MSE}$ & $E_{RMSE}$ & $A_c$ \\
\hline
Geolife Trajectories\cite{ref7}& $0.012 866$& $0.000 108$&$0.010 221$&$0.991 8$  \\
\hline
\makecell{T-drive Taxi\\Trajectories\cite{ref12}}& $0.014222$& $0.000 130$&$0.011 289$&$0.950 2$  \\
\hline
ETData\cite{ref14}& $0.005 407$& $0.000 020$&$0.004 319$&$0.988 9$  \\
\hline
\end{tabular}}
\label{tab1}
\end{center}
\end{table} 
It is observed from the Fig. \ref{fig:galaxy} and Table \ref{tab1} that the prediction results are highly accurate, showing low error values on all error metrics with good results.
\begin{figure*}
	\centering
	\begin{minipage}[t]{0.33\linewidth}
		\centering
		\includegraphics[width=2.6in]{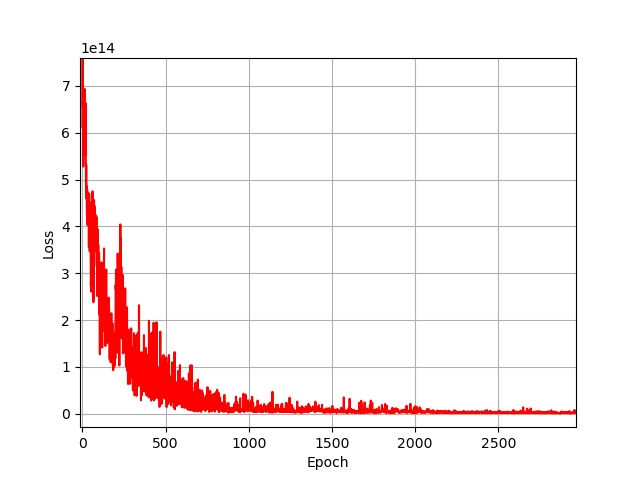}
		\caption{Loss of TPPD algorithm}
		\label{figure4}
	\end{minipage}%
	\begin{minipage}[t]{0.33\linewidth}
		\centering
		\includegraphics[width=2.6in]{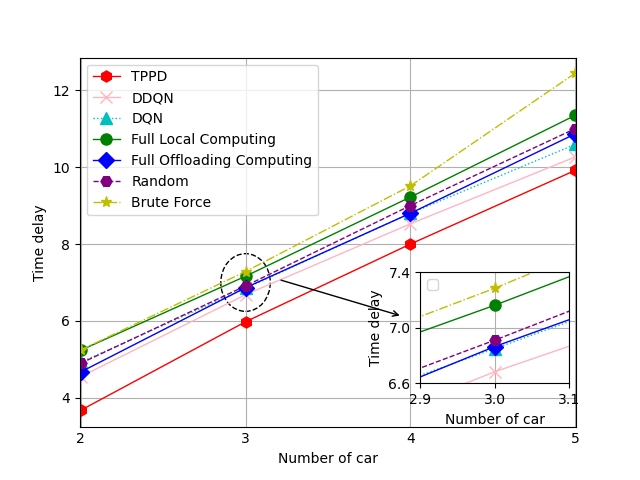}
		\caption{Comparison of overall time consumption of various algorithms}
		\label{figure5}
	\end{minipage}
	\begin{minipage}[t]{0.33\linewidth}
		\centering
		\includegraphics[width=2.2in]{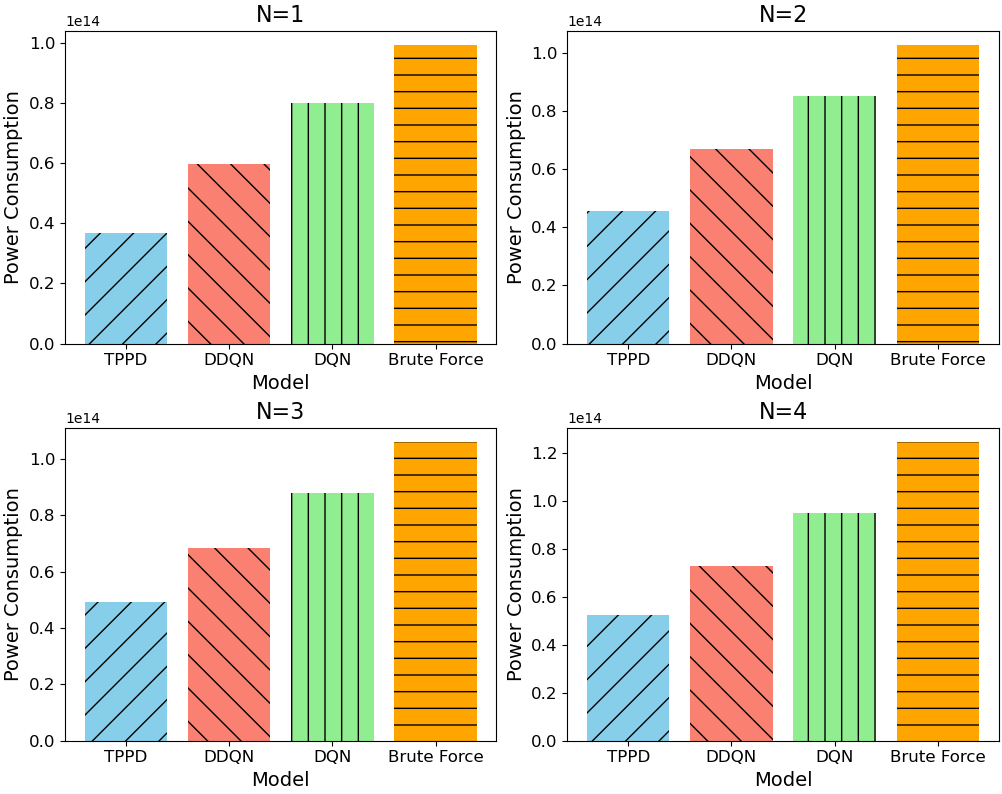}
		\caption{Comparison of power consumption with different number of vehicles}
		\label{figure6}
	\end{minipage}
 \end{figure*}
\subsection{Performance of the TPPD Algorithm}
Fig. \ref{figure4} shows the loss of the TPPD algorithm. It can be seen that as the number of iterations increases, the overall loss decreases and the network converges. There are many schemes for task offloading. In this paper, all offloading, all local processing, randomized policy, DDQN, DQN, and traditional exhaustive algorithms are selected as references. DDQN, DQN, and exhaustive algorithms involve making a decision at the present moment, so the overall latency also includes the time to get the policy, expressed as \(Time_{\text{delay}} = C + Time_{\text{decision}} \times \psi\), where \(C\) denotes the time when the vehicle task is unloaded and completed ,\(\psi\) denotes the penalty factor, with a value range of \([0, 1]\). The longer the time taken to obtain the strategy, the larger the value of \(\psi\). As shown in Fig. \ref{figure5}, the horizontal axis represents the number of different end devices and the vertical axis represents the overall delay cost. The performance of our proposed algorithm is optimal in terms of overall efficiency, and the time consumed is relatively minimal. The power consumption of the whole system is calculated by \(p_{\text{cpu}} = Time_{\text{delay}} \times J\), where \(J\) indicates the power of the base station. The comparison of TPPD, DDQN, DQN, and exhaustive algorithms is shown in Fig. \ref{figure6}. The four subgraphs indicate the differences in the number of cars. The graph shows that our algorithm consumes the least of energy.
\section{Conclusion}
In this paper, we presented a computation pre-offloading strategy for on-board tasks in multi-vehicle wireless networks with multiple MEC servers. Firstly, vehicle trajectory prediction was carried out based on their historical trajectory analysis, which was leveraged for non-detachable task offloading analysis. Then,  we designed a task offloading algorithm based on DDQN missioned to analyze pre-offloading strategies. Our experimental results showcased that our
proposed TPPD algorithm can accurately predict the trajectory of vehicles and derive a good offloading strategy, which can achieve the initial goal and obtain the offloading strategy more efficiently.

\end{document}